\documentclass[10pt,a4paper]{article}
\usepackage{epsfig}
\usepackage{amsmath}
\usepackage{amssymb}
\usepackage{color}
\usepackage{rotating}
\usepackage{hyperref}
\usepackage{tikz}
\usetikzlibrary{arrows,shapes,decorations,positioning}
\tikzstyle{line} = [draw, -latex']
\tikzstyle{con} = [thick, dashed, draw, angle 90 reversed-angle 90 reversed]

\usepackage{todonotes}
\usepackage{mathtools}
\usepackage{xfrac}
\usepackage{nicefrac}

\renewcommand{\AA}{\protect\hbox{$\overset{_{\circ}}{\text{A}}$}}
\newcommand{\Ha}{H$\alpha$}

\newcommand{\euler}{\text{\Large\it e}}

\def\aj{AJ}                 % The Astrophysical Journal 
\def\apj{ApJ}                 % Astrophysical Journal
\def\apjl{ApJL}                 % Astrophysical Journal Letters
\def\apjs{ApJS}               % Astrophysical Journal, Supplement
\def\mnras{MNRAS}             % Monthly Notices of the RAS
\def\aap{A\&A}                 % Astronomy & Astrophysics
\begin{document}

\title{\bf Hierarchical Bayesian approach for estimating physical properties in spiral galaxies: \\ Age Maps for M74}

\author{M. Carmen S\'anchez Gil$^{1,3}$, Angel Berihuete$^{1,2}$, Emilio J. Alfaro$^{3}$, \\
Enrique P\'erez$^{3}$ and Luis M. Sarro$^{2,4}$}

\date{}
\maketitle
%\address{
\begin{flushleft}
{\small
$^{1}$ Dpt. Statistics and Operations Research, UCA,
Avd. Universidad s/n, Jerez de la Fra., Spain \\
$^2$ Spanish Virtual Observatory \\
$^{3}$ Instituto de Astrof\'{i}sica de Andaluc\'ia (CSIC), E18008, Granada, Spain\\
$^{4}$ Dpt. Inteligencia Artificial, ETSI Inform\'atica, UNED, Juan del Rosal, 
6, E-28040, Madrid, Spain
%Production Editor, \jpcs, \iopp, Dirac House, Temple Back, Bristol BS1~6BE, UK}
}
\end{flushleft}

%\ead{mcarmen.sanchez@uca.es}

\begin{abstract}
 One of the fundamental goals of modern Astronomy is to estimate the physical parameters of galaxies from images in different spectral bands. We present a hierarchical Bayesian model for obtaining age maps from images in the \Ha\ line (taken with Taurus Tunable Filter (TTF)), ultraviolet band (far UV or FUV, from GALEX) and infrared bands (24, 70 and 160 microns ($\mu$m), from Spitzer). As shown in  \cite{2011MNRAS.415..753S}, %(DOI: 10.1111/j.1365-2966.2011.18759.x),
 we present the burst ages for young stellar populations in the nearby and nearly face on galaxy M74. 
 
As it is shown in the previous work, the \Ha\ to FUV flux ratio gives a good relative indicator of very recent star formation history (SFH). As a nascent star-forming region evolves, the \Ha\ line emission declines earlier than the UV continuum, leading to a decrease in the \Ha\/FUV ratio. Through a specific star-forming galaxy model (Starburst 99, SB99), we can obtain the corresponding theoretical ratio \Ha\ / FUV to compare with our observed flux ratios, and thus to estimate the ages of the observed regions. 

Due to the nature of the problem, it is necessary to propose a model of high complexity to take into account the mean uncertainties, and the interrelationship between parameters when the \Ha\ / FUV flux ratio mentioned above is obtained. 
To address the complexity of the model, we propose a Bayesian hierarchical model, where a joint probability distribution is defined to determine the parameters (age, metallicity, IMF), from the observed data, in this case the observed flux ratios \Ha\ / FUV. The joint distribution of the parameters is described through an i.i.d. (independent and identically distributed random variables), generated through MCMC (Markov Chain Monte Carlo) techniques.
\end{abstract}

%\keywords{{spectroscopy ---  galaxies: star formation --- Bayesian estimation: Hierarchical model}}

%%%%%%%%%%%%%%%%%%%%%%%%%%%%%%%%%%%%%%%%%%%%%%%%%%
\section{Introduction}

The study of the star formation history (SFH) and star formation rate (SFR) in galaxies provide vital information on the evolutionary properties of galaxies and the physical processes which drive that evolution.
The variation in star formation (SF) activity across galaxies of different types is influenced by many factors, including gas content, mass and dynamical environment (\cite{Kennicutt1998}). 
But even galaxies with the same morphological type show a diversity in SF and enrichment  histories (e.g.  \cite{Grebel2000}). Some SFR measurements using \Ha\ emission suggest episodic bursts of SF, rather than continuous SF ( \cite{Glazebrook1999}).  Trends in SFR and SFH across galaxy types allow predictions of galactic evolution with cosmic lookback time.  
In those galaxies with resolved stars it is possible to determine the SFH from analysis of the color-magnitude diagrams in regions with enough stars to reach sensible results. However individual stars are only resolved in the closest galaxies, so global SF properties of galaxies are mostly obtained by integrated light measurements. The most common diagnostic methods use measurements in the ultra-violet (UV), far-infrared (FIR), or nebular recombination lines (\cite{Kennicutt1998}). Of particular interest for this work are measurements in the UV and the \Ha\ recombination line.

One way to deal with the study of the SFH in galaxies is to obtain the age map of the galaxy. 
In this paper we propose to study the spatial distribution of SFH in galaxies on a pixel by pixel basis, without any a priori hypothesis about the extent and spatial delineation of the star forming regions. Ages are obtained from comparison between \Ha\ and UV emission, so this method is only sensitive to exploring the age map of the youngest stellar population, less than about 10 Myr. 
This age map should provide a global vision of the current star formation processes taking place in the galactic disk, the maximum scale of coherent star formation, and its relation with other large scale processes such as density waves. 

This work is part of a larger study of a sample of nearby galaxies, where star forming regions are spatially resolved, in order to place the relationship between star formation, ultraviolet and \Ha\ emission on a stronger empirical foundation ( \cite{2011MNRAS.415..753S}; hereinafter Paper I). 

The comparison of UV and \Ha\ fluxes is used as a tracer of the recent star formation history. While the infrared images are used to correct for extinction via the TIR to FUV ratio. This flux ratio is defined as a robust and universal tracer of dust extinction ( \cite{BuatXu1996};  \cite{Buat1999};  \cite{Gordon2000}). 
The measured \Ha/FUV ratios are compared with the stellar population synthesis (SPS) models to study the dynamics of stellar populations of different ages within the galaxies; we use Starburst99, SB99 ( \cite{Leitherer1999};  \cite{VazquezLeitherer2005})\footnote{http://www.stsci.edu/science/starburst99/}.
Model flux ratios are compared to the observed values for each pixel
in the images, resulting in two dimensional spatial {\it age maps}.

We address the problem of deriving the {galaxy age map} 
from {\Ha/FUV} flux ratio images by establishing a {probabilistic framework}
which explain the relationship of the random variables involved in the problem. 

 This relationship will be formulated in terms of a {joint probability distribution} 
given the observations and their uncertainties. 
 We define this joint probability distribution in terms of a {hierarchical Bayesian model (HBM)} and use Markov Change  Monte Carlo (MCMC) to describe it ({Nested Sampling})

 In order to set up the HBM we need to generate a \emph{model flux 
ratio}~\footnote{We prefer use the term \emph{model} 
instead than \emph{true} because its value depends on the assumption that
SB99 is an appropriate model} from SB99 model 
and compare it with the observations by 
using a {likelihood}. 
 Finally, we explore the parameters of interest (age) marginalizing nuisance parameters in the {posterior distribution} defined by HBM model.

%%%%%%%%%%%%%%%%%%%%%%%%%%%%%%%%%%%%%%%%%%%%
\section{Observations}\label{sec:observations}

\begin{figure*}[t]
	\centering
		\includegraphics[width=0.8\textwidth]{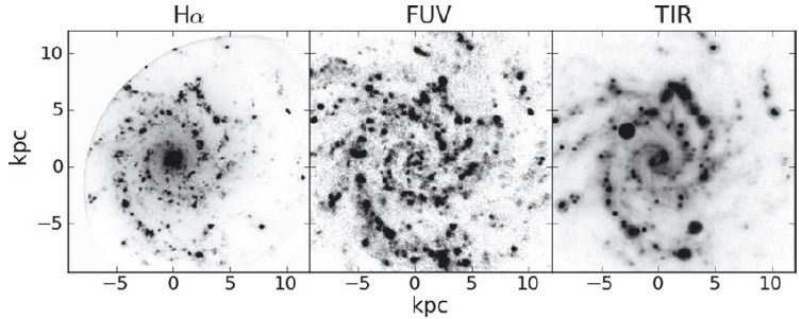}
	\caption{Examples of processed frames in  \Ha\ (left), far ultraviolet (FUV; centre), and total infrared (TIR; right) for the galaxy M74. 
The images have been resampled to have identical size, orientation, and pixel scale (1.5''/pix).}\label{fig1}
\end{figure*}

The \Ha\ image was taken with the Taurus Tunable Filter (TTF;
 \cite{BlandHawthorn1997}) on the William Herschel Telescope (WHT) on 1999 March 4$-$6.
Conditions were photometric with stable seeing of 1.0 arcsec.  TTF was tuned to a bandpass
of width $\Delta \lambda = 20 $\,\AA\  centred at $\lambda_{\rm c} = 6570$\,\AA. The intermediate-width 
R0 blocking filter ($\lambda_{\rm c}/\Delta\lambda = 6680/210$\,\AA) was used to remove
transmissions from all but a single interference order. The pixel scale is 0.56 arcsec.
More details on the data and data reduction can be found in Paper I. 

The UV image comes from the Nearby Galaxies Survey of the Galaxy Evolution Explorer mission
(NGS survey, GALEX,  \cite{Martin2005}). This survey contains well-resolved imaging (1.5 arcsec pix) of 296 and 
433 nearby galaxies for GR2/GR3 and GR4 releases, respectively, in two passbands: a narrower far-ultraviolet 
band (FUV; $\lambda_{\rm eff}/\Delta \lambda=1516 / 268$\,\AA), and a broader near-ultraviolet band 
(NUV; $\lambda_{\rm eff} = 2267 / 732$\,\AA). 

Archival {\em Spitzer}\footnote{http://irsa.ipac.caltech.edu/data/SPITZER/SINGS/summary.html} images, at 24, 70, and 160$\mu$, were used 
to provide additional estimates of extinction. 
They were resampled to a common 1.5 arcsec/px scale and combined into an image of TIR flux, 
$F_{TIR} = \zeta_1 \nu F_{\nu}(24 \mu)+ \zeta_2 \nu F_{\nu}(70 \mu)+\zeta_3 \nu F_{\nu}(160 \mu)$, 
with [$\zeta_1$,$\zeta_2$,$\zeta_2$] = [1.559,0.7686,1.347] (Dale \& Helou 2002).  

The extinction within  our galaxy is corrected using the   \cite{Schlegel1998}\footnote{http://www.astro.princeton.edu/$\sim$schlegel/dust/data/data.html} dust maps, which measure the Galactic extinction in all directions. 
The internal reddening is corrected using a straight relation between the A$_{FUV}$ extinction factor and the $F_{TIR}/F_{FUV}$ flux ratio:  $A_{FUV} = -0.0333 y^3 + 0.3522 y^2+1.1960 y + 0.4967$, where $y=log(F_{dust}/F_{FUV})$ ( \cite{Buat2005}). 
The $F_{TIR}/F_{FUV}$ ratio is a robust and universal tracer of dust extinction, almost independent of the dust/stars geometry and of the dust properties, provided that the galaxies are actively forming stars ( \cite{BuatXu1996};  \cite{Buat1999};  \cite{Gordon2000}). 
The A(H$\alpha$) extinction factor was derived through the relation $A_{FUV} = 1.4 A(H \alpha)$ ( \cite{Boissier2005}).

The resampled and aligned \Ha, FUV and TIR images are shown in Fig.\ref{fig1}, center panel.

%%%%%%%%%%%%%%%%%%%%%%%%%%%%%%%%%%%%%%%%%%%%%
\section{Methodology} \label{sec:methodology}

In this section, we address the problem of deriving the galaxy age map 
(GAM) from $F_{H_{\alpha}} / f_{FUV}$ flux ratio images by establishing a probabilistic framework 
which explain the relationship between random variables involved in the problem. 
This relationship will be formulated in terms of a joint probability distribution 
given the observations and their uncertainties using a hierarchical Bayesian model 
(HBM) \cite{Gelman03}. Specifically, we want to 
describe the probability distribution $p(\mbox{age} | r_{obs})$, where $r_{obs}$ is the observed \Ha/FUV flux ratio.

Because the use of SB99 we extend the previous distribution to 
\begin{equation}\label{eq:posteriorDistribution}
p(\theta | r_{obs}), \quad \theta \in \mathbb{R}^3,
\end{equation}

where the first component of $\theta$ is related 
to the age, the second is related to the initial mass 
function (IMF) and third component is related to metallicity. Furthermore, by 
marginalizing we can derive the age distribution according to 

\begin{equation}\label{eq:marginalizationPost}
p(\mbox{age} | r_{obs}) = p(\theta_1 | r_{obs}) = \iint p(\theta | r_{obs}) \, d 
\theta_2 d \theta_3.
\end{equation}
The joint probability distribution $p(\theta | r_{obs})$ can be rewritten by using
the Bayes theorem according to 
\begin{equation}\label{eq:bayesdescomposition}
p(\theta | r_{obs}) = \frac{p(r_{obs} | \theta) p(\theta)}{p(r_{obs})} \propto  p(r_{obs} | \theta) p(\theta),
\end{equation}
where $p(\theta)$ is the prior distribution and $p(r_{obs} | \theta)$ is
the likelihood function and $p(r_{obs})$ is a normalization constant 
according to 
\begin{equation}\label{eq:evidence}
p(r_{obs}) = \int p(r_{obs} | \theta) p(\theta) \, d\theta
\end{equation}

\subsection{The prior distribution}\label{sub:priors}

\begin{figure}
	\centering
	\includegraphics[width=0.75\textwidth]{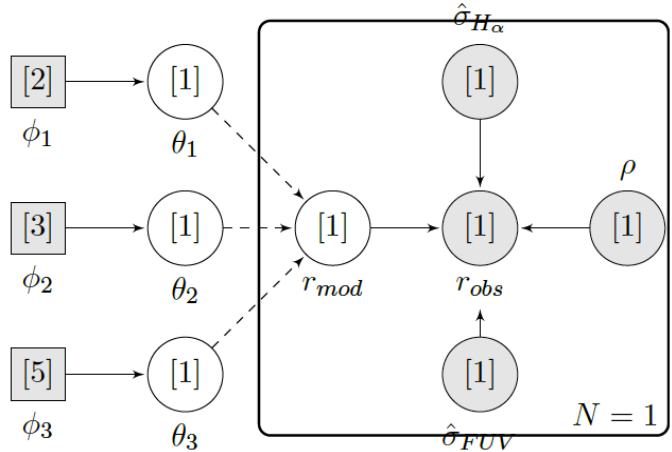}
	\caption{Hierarchical model using plate notation. Hyperparameters and
parameters are written with $\phi_k$ and $\theta_k$ respectively. 
Observations and model ratios are inside the plate. The letter $N$ 
notes the sample size per pixel, in our case, we only observed
one pixel at time, then $N=1$. Squares are fixed quantities and circles
random variables. If the node is shaded then the variable is known (observed).}
	\label{fig:Hierarchicalmodel}
\end{figure}

The HBM is plot in figure~\ref{fig:Hierarchicalmodel} using plate notation. Prior 
distributions are represented with solid line linking hyperparameters $\phi_i$ and 
$\theta_i$ with $i \in \{1,2,3\}$. These hypermaramters are konwn and fixed in order 
to draw the parameters
$\theta$ according to different prior distributions. Priors for $\theta$ were set as
\begin{eqnarray}
\theta_1 & \sim & \mathcal{U}(0,10)  \nonumber\\ 
\theta_2 & \sim & \mbox{Cat}(3) \\ 
\theta_3 &\sim  &   \mbox{Cat}(5) \nonumber
\end{eqnarray}
where $\mbox{Cat}(n)$ is a categorical distribution with $n$ categories. For the IMF, 
the categories are: first, a Salpeter law with $\alpha=2.35$ and $M_{up} = 
100M_{\odot}$; second, a truncated Salpeter law with $\alpha=2.35$ and $M_{up} = 
30M_{\odot}$; third, and a Miller-Scalo law with $\alpha=3.3$ and $M_{up} = 
100M_{\odot}$. For metallicity, categories are $Z = $ 0.04, 0.02 (solar, 
$Z_{\odot}$), 0.008, 0.004 and 0.001. Finally the prior probability distribution for 
age was set to the uniform distribution in the interval 0-10 Myr.

\subsection{The likelihood}\label{sub:truefluxratio}
\begin{figure}[htbp]
	\centering
		\includegraphics[width=0.75\textwidth]{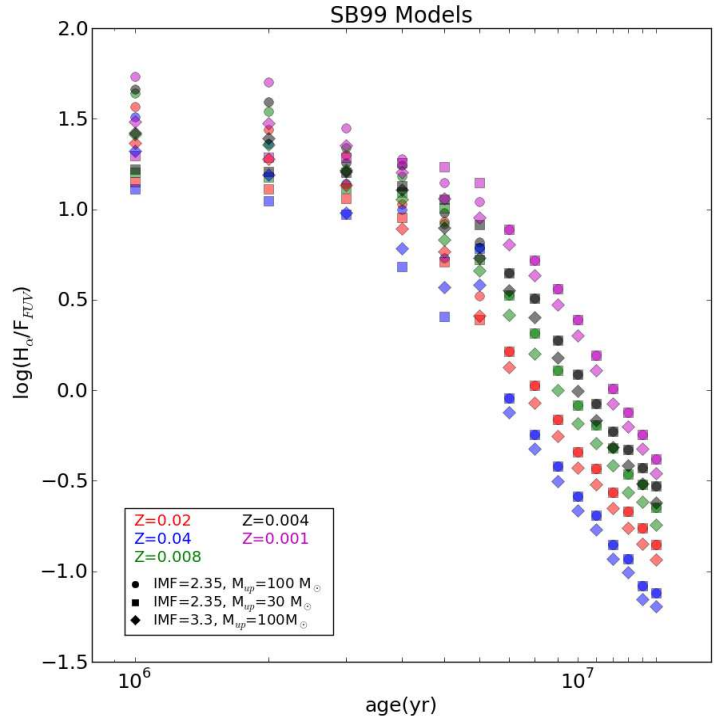}
	\caption{Model flux ratios for every combination of SB99 parameters.}
	\label{fig:ModelosSB99}
\end{figure}

The likelihood function is represented by lines linking the variables $\theta_i$
with $r_{obs}$ in Fig.~\ref{fig:Hierarchicalmodel}. In order to obtain \emph{model 
flux ratio}~\footnote{We prefer use the term \emph{model} 
instead than \emph{true} because its value depends on the assumption that
SB99 is an appropriate model} $r_{mod} = F_{H_{\alpha}} / f_{FUV}$ from SB99 model, we need to 
set up SB99 with parameter $\theta = (\theta_1, \theta_2, \theta_3)$, but
let us notice that the SB99 model only permits a finite set of values for 
$\theta_2$ and $\theta_3$, see Fig.~\ref{fig:ModelosSB99}. This situation means we 
cannot work with a continuous parameter space in three dimensions but a grid of 
parameters $\{ \theta \}^{\mbox{grid}}$. 

Because this finite combination of parameters in SB99, we can only derive
a finite number of model flux ratios from them.
Therefore, in order to obtain more model flux ratios different from those on the grid, 
we use an Artificial Neural Network (ANN) which will interpolate the grid of 
parameters. A deep information about ANN can be found in \cite{Haykin1999}. 
Specifically the ANN is a 
\emph{multilayer perceptron} with four 
layers: two hidden layers with 20 and 50 nodes, the input layer for the parameters 
$\theta$ and the output layer for the \Ha\ and FUV flux. The ANN was trained
with 70~\% and validate with 30~\% of the grid points. The uncertainties due this interpolation were considered negligible. Dashed lines in Fig.~\ref{fig:Hierarchicalmodel} represent the ANN.

\subsubsection{Independence between pixels}\label{subsub:CASE_A}%
For explanation purposes this section assume independence between the fluxes of
different pixels. Also the results are established for an unique pixel with $r_{obs} > 0 $. Once the SB99 have generated the model fluxes we assume that the observed flux from each pixel has a normal distribution
\begin{eqnarray}
F_{H_{\alpha}, obs} = F_{H_{\alpha}, mod} + \epsilon_{H_{\alpha}}, & \epsilon_{H_{\alpha}} \sim \mathcal{N}(0,\hat{\sigma}_{H_{\alpha}}) \\
f_{FUV, obs} = f_{FUV, mod} + \epsilon_{FUV}, & \epsilon_{FUV} \sim \mathcal{N}(0,\hat{\sigma}_{FUV}),
\end{eqnarray}
where $F_{H_{\alpha}, obs}$, $f_{FUV, obs}$, $F_{H_{\alpha}, mod}$ and 
$f_{FUV, mod}$ are the observed and model \Ha\ and FUV fluxes for the selected pixel 
respectively. The quantities $\hat{\sigma}_{H_{\alpha}}$ and 
$\hat{\sigma}_{H_{\alpha}, obs}$ are the flux uncertainties due to the instrumental, 
reduction and correction processes. Therefore, $r_{obs} = F_{H_{\alpha}, obs} / 
f_{FUV, obs}$ is the ratio of two correlated normal random variables, which exact 
distribution is given by ( \cite{Hinkley1969})

\begin{align}
\psi(r) &=  \frac{b(r)d(r)}{\sqrt{2\pi}\hat{\sigma}_{H_{\alpha}} \hat{\sigma}_{FUV}}
\left[ \Phi\left( \frac{b(r)}{\sqrt{1-\rho^2}a(r)}\right) - 
\Phi\left( -\frac{b(r)}{\sqrt{1-\rho^2}a(r)}\right) \right]%+
\nonumber \\
& + \frac{\sqrt{1-\rho^2}}{\pi\hat{\sigma}_{H_{\alpha}}\sigma_{FUV}a^2(r)}\exp\left\{-\frac{c}{2(1-\rho^2)}\right\}
\label{eq:fddR}
\end{align}
where,
\begin{align}
a(r) &= \,\,\, \left( \frac{r^2}{\hat{\sigma}^2_{H_{\alpha}}}-\frac{2\rho r}{\hat{\sigma}_{H_{\alpha}} \hat{\sigma}_{FUV}}+
\frac{1}{\hat{\sigma}^2_{FUV}}\right)^{1/2}
\\
b(r) &= \,\,\, \frac{r^2}{\hat{\sigma}^2_{H\alpha}}-\frac{2\rho r}{\hat{\sigma}_{H\alpha} \hat{\sigma}_{FUV}}+
\frac{1}{\hat{\sigma}^2_{FUV}}
\\
c \,\,\, &= \,\,\, \frac{F^2_{H_{\alpha}, mod}}{\hat{\sigma}^2_{H\alpha}}-\frac{2\rho F_{H_{\alpha}, mod} f_{FUV, mod} }{\hat{\sigma}_{H\alpha}\hat{\sigma}_{FUV}}+
\frac{f^2_{FUV, mod}}{\hat{\sigma}^2_{FUV}}
\\
d(r) &= \,\,\, \exp\left\{\frac{b^2(r)-ca^2(r)}{2(1-\rho^2)a^2(r)}\right\}
\end{align}
where $\rho$ is the correlation between \Ha\ and FUV, and $\Phi$ is the cumulative 
density distribution of the standard normal,
\begin{equation}
\Phi(z) = \int_{-\infty}^y \phi(u)du, \quad \phi(u) = \frac{1}{\sqrt{2\pi}}\euler^{-\frac{1}{2}u^2}
\end{equation}

The ${F}_{H\alpha}$ and ${f}_{FUV}$ are correlated, in fact in normal star forming  galaxies it is expected a constant flux ratio
$\frac{{F}_{H\alpha}}{{f}_{FUV}} \simeq 11$ \AA\ 
(  \cite{Meurer2009},  \cite{Iglesias2004}).

In order to compare model and observed flux ratio, pixel-by-pixel, we consider the likelihood according to
\begin{equation}\label{eq:Likelihood}
p(r_{obs} | \theta ) = \psi (r_{obs} | F_{H_{\alpha}, mod},
f_{FUV, mod}, \hat{\sigma}_{H\alpha},\hat{\sigma}_{FUV}, \rho)
\end{equation}
where $F_{H_{\alpha}, mod}$ and $f_{FUV, mod}$ are obtained from $\theta$ through the  ANN. 
The theoretical correlation, $\rho$, between $F_{H\alpha}$ and $f_{FUV}$ is estimated from all the pixels of the images .

\subsection{The posterior distribution}
We characterize the posterior distribution in equation 
\ref{eq:posteriorDistribution} by extracting independent samples
using the Nested Sampling Algorithm (NSA) \cite{skilling2006}. The aim of
the NSA is the estimation of $p(r_{obs})$ defined in~\ref{eq:evidence}, and
as by-product an independent sample of the posterior distribution is obtained. The 
NSA is based on the relationship between the likelihood $p(r_{obs}|\theta)$ and the 
\emph{prior volume}
$X(\lambda)$ defined by
\begin{equation}\label{eq:priorvolume}
X(\lambda) = \int_{p(r_{obs}|\theta) > \lambda} p(\theta) \, d\theta,
\end{equation}
which means the bulk of prior distribution contained within an iso-contour of
the likelihood.
 
Finally we obtain the posterior distribution of our parameter of interest (age) by 
marginalizing nuisance parameters in the posterior distribution as we set in 
equation~\ref{eq:marginalizationPost}.

\begin{figure}[t]
%\hspace*{-1.3cm}
 \includegraphics[width=\textwidth]{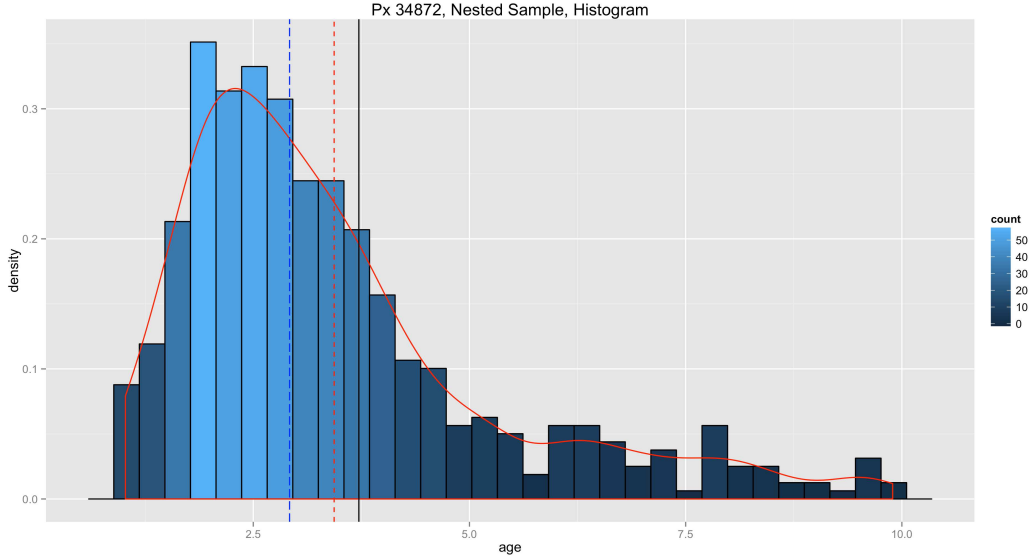}
   \caption{Sample of the marginal probability distribution for the age parameter, for the pixel \#34872 of the image. The blue and red dashed lines represent the median and the mean respectively. The solid black line shows the age with the maximum likelihood.}
 \label{fig:AgeDensity}
 \end{figure}
 
Figure \ref{fig:AgeDensity} shows the sample of the marginalized the age distribution for a certain pixel of the image, as an example of the resulting posterior distribution.

%%%%%%%%%%%%%%%%%%%%%%

\begin{figure*}[!h]
\centering
%\hspace*{-1.2cm}
 \includegraphics[width=0.85\textwidth]{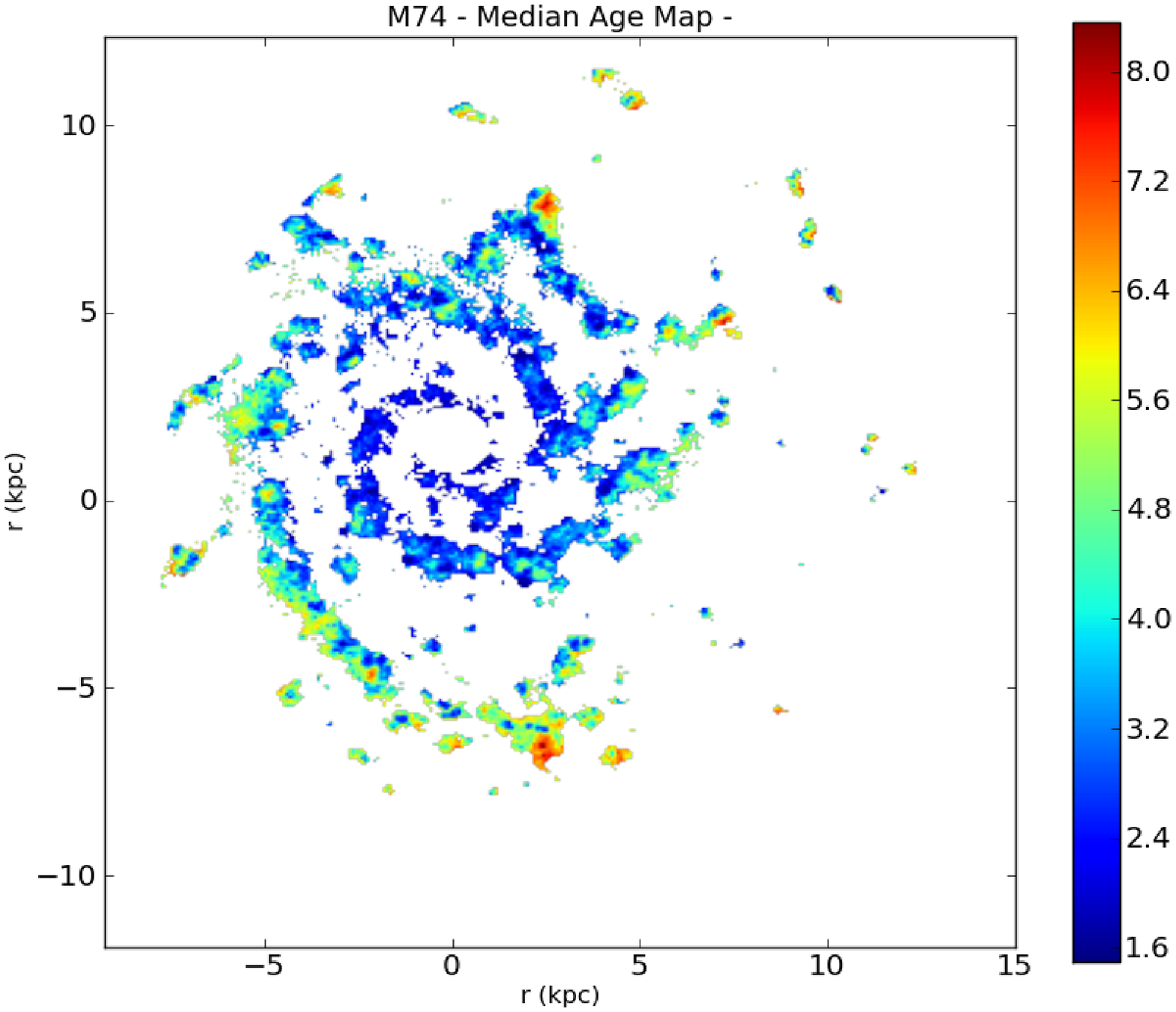}
% \hspace*{-1.2cm}
 \includegraphics[width=0.85\textwidth]{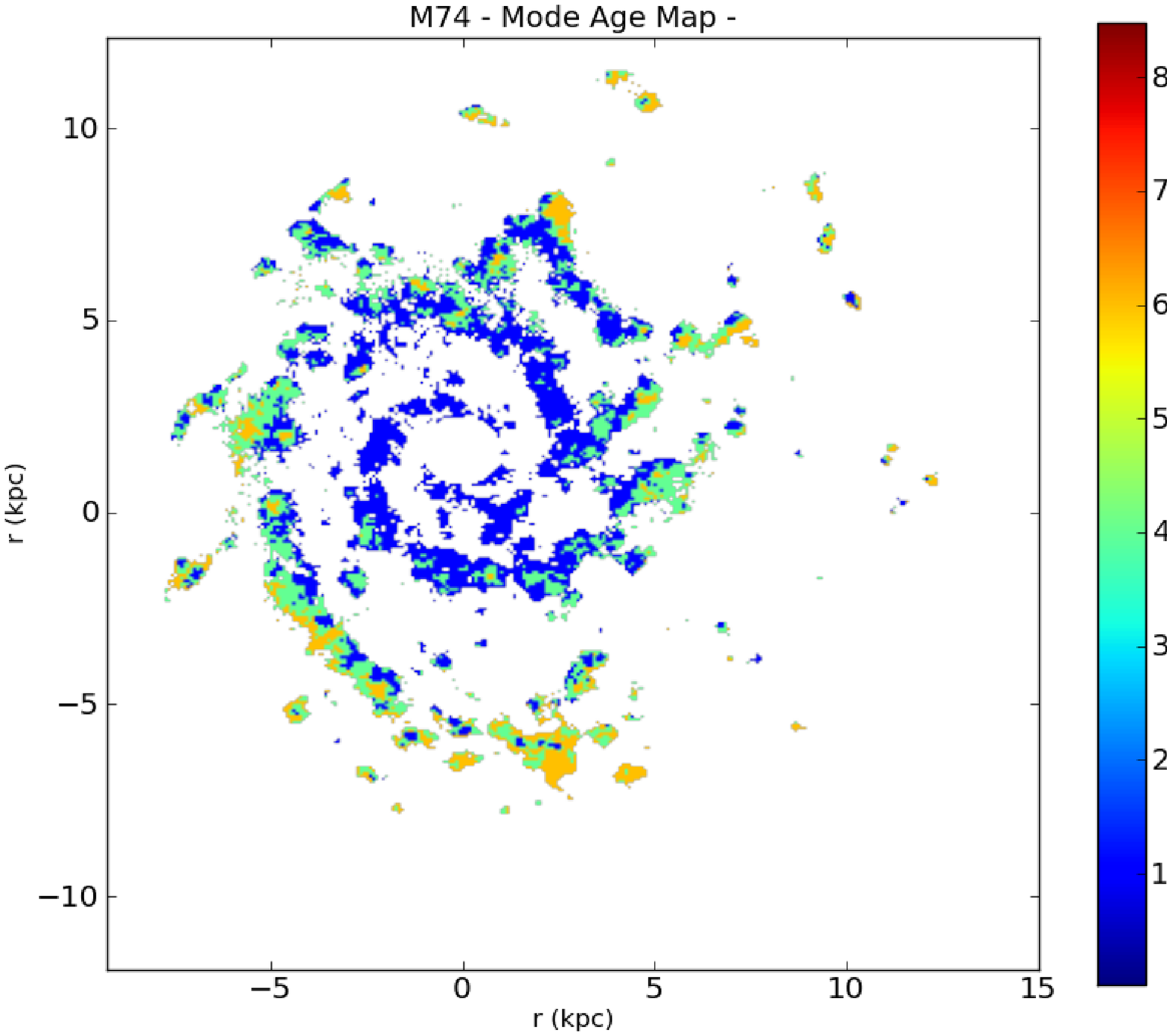}
   \caption{Comparison between the Agemaps for M74: on the top when applying the robust, astrophysical method in Paper I. And on the bottom, applying the Bayesian approach.}
 \label{fig:Agemaps1}
 \end{figure*}

%%%%%%%%%%%%%%%%%%%%%%%%%%%%%%%%%%%%
\section{Results}\label{sec:results}

\begin{figure*}[!h]
\centering
%\hspace*{-1.15cm}
 \includegraphics[width=0.6\textwidth]{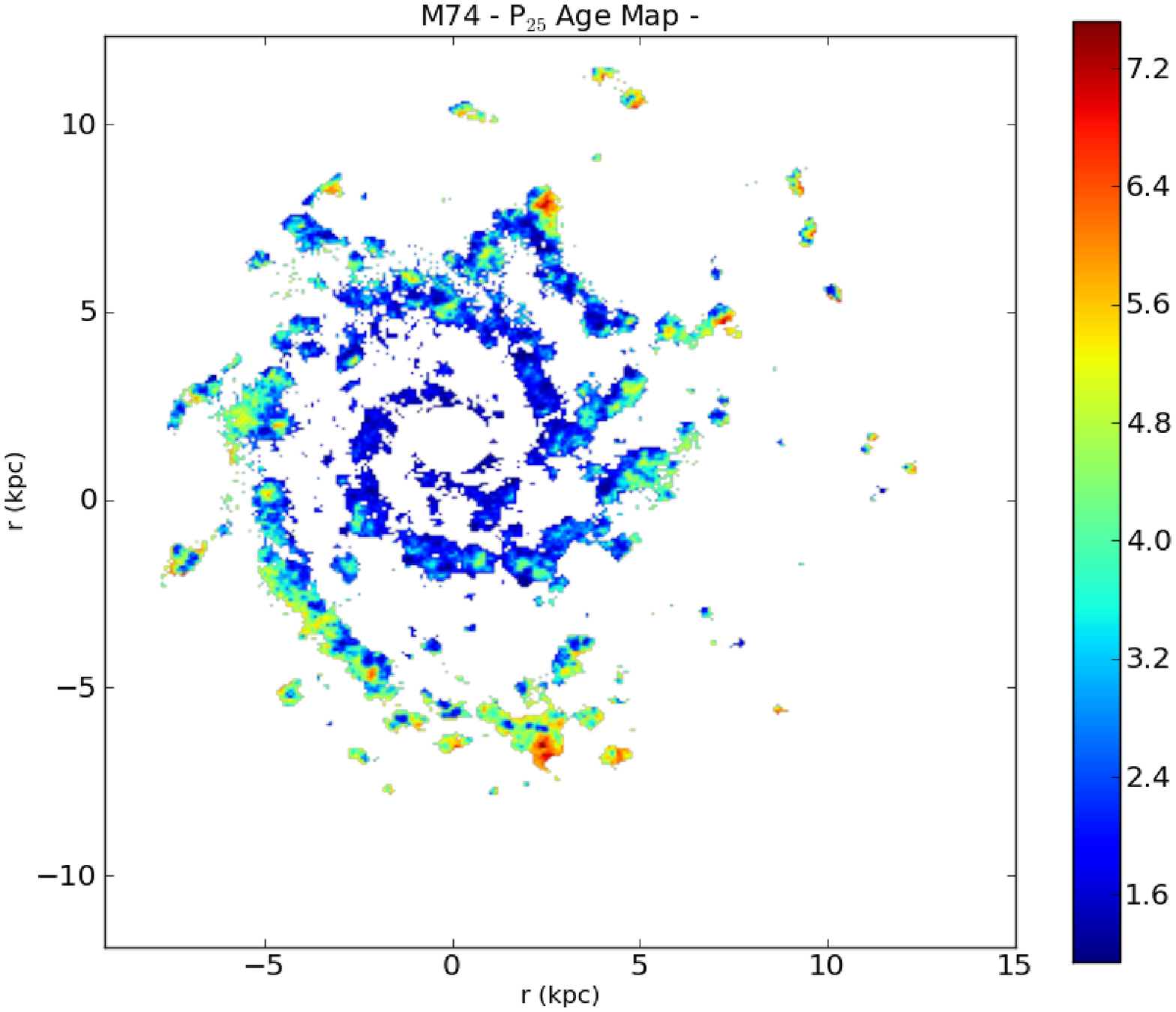}
% \hspace*{-1.4cm}
 \includegraphics[width=0.6\textwidth]{M74_bayes.eps}
 % \hspace*{-1.4cm}
  \includegraphics[width=0.6\textwidth]{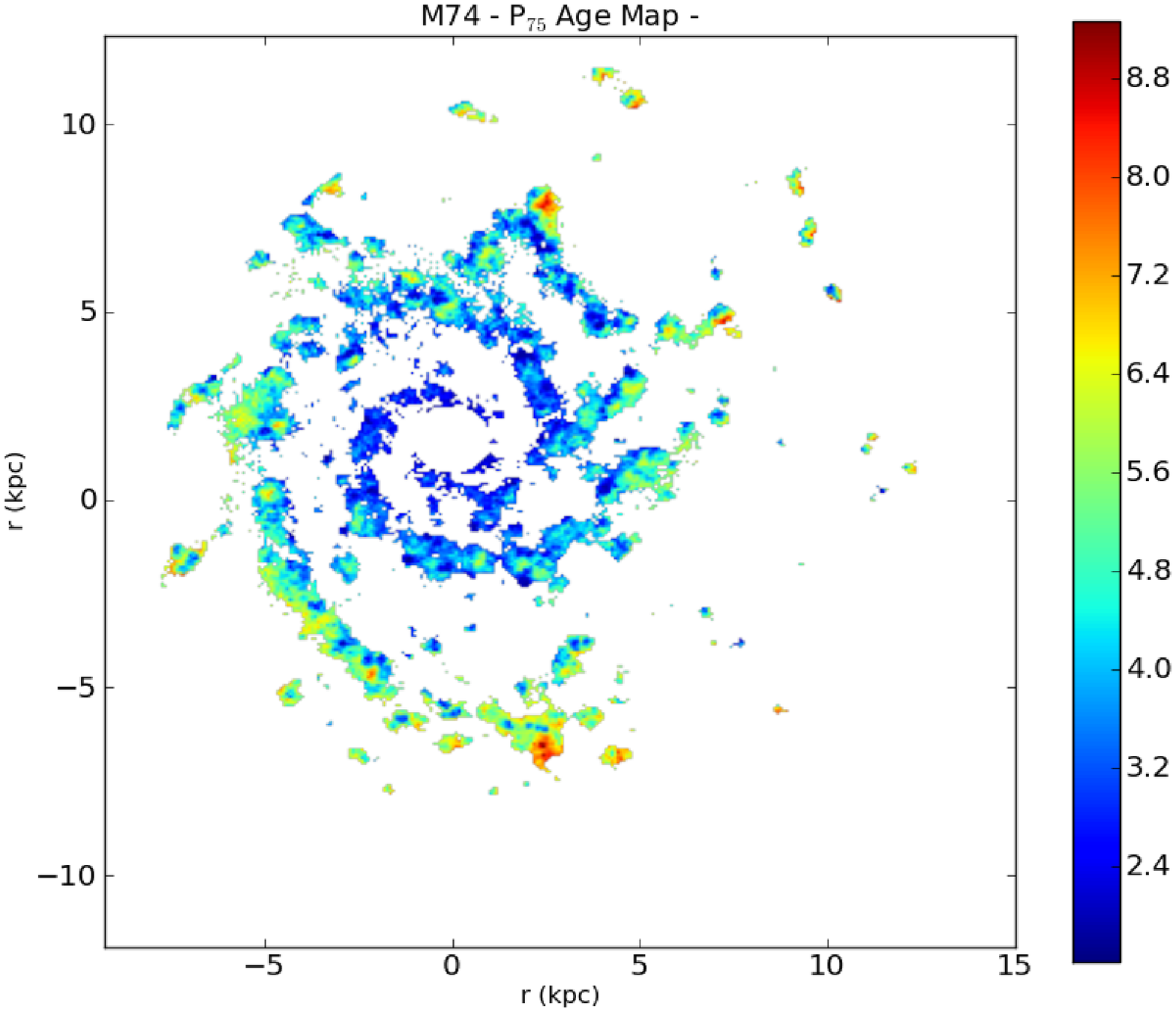}
   \caption{Agemaps for M74: On the top it is shown the Percentile 25, $P_{25}$, of the Age distribution. The central age map corresponds to the Median ages, and on the bottom the $P_{75}$. With this comparison we can check the robust of this methodology for age estimation.}
 \label{fig:Agemaps2}
 \end{figure*}

Figures \ref{fig:Agemaps1} and \ref{fig:Agemaps2} show some age maps for M74. 
In the Figure \ref{fig:Agemaps1}, we can compare between the age maps for M74 obtained in Paper I, by applying a robust, only based on astrophysical models. Whereas on the left, it is shown the age map when applying the Bayesian approach. The main difference, and more remarkable is that the first one is a discrete age map mode. However, with the Bayesian inference we can determine a single age to each pixel. Resulting therefore in more rich age patterns ans structures, which could not been observed with the latter. 

To have an idea of the uncertainty in the age estimation, it is included the Figure   \ref{fig:Agemaps2}, where it is represented the 50\% central distribution of the age.
On the left it is shown the Percentile 25, $P_{25}$, of the age distribution. The central age map corresponds to the median ages, $P_{50}$, and on the left the $P_{75}$. With this comparison  we get as a confidence interval, checking out the robust of this methodology for age estimation. 

Finally, with respect to the physical conclusions we can observe that these are basically the same that were obtained in Paper I. 
The age map shows an age gradient from the inner to the outer parts of the galaxy, from very recent to less recent episodes of star formation, in agreement with previous authors ( \cite{Cornett1994},  \cite{Roy2000}).
 
Specifically, we find that the \Ha\ luminosity decrease in radius is more pronounced in the inner 5 to 6~kpc, while the UV luminosity
shows a shallower rate of change. Consequently, the \Ha/FUV ratio decreases with radius indicating an age increases in 
the outward direction.    

On more localised scales, the short arm that opens S-SW at 4~kpc shows a clear age gradient across it. 
However, the outer longer arm, that runs SE-S at $\sim 5$ to 10~kpc, showed a less marked age gradient across its width in Paper I. But improving the "resolution" of our age maps with the bayesian technique, it is more remarkable the age gradient across the outer longer arm as well.   
If the age gradients across spiral arms are a direct product of the spiral density wave, then the dilution of the 
gradient in this southern arm may be related to the weakness of the density wave or the approach to corotation.
As discussed by  \cite{Efremov2009}, the presence of a shock produced by the spiral density wave, (made visible by dust 
lanes along the spiral arms), is incompatible with the creation of star forming complexes, because of the absence
of visible dust lanes in this arm, despite a chain of complexes along its length. The thickness of the longer arm is basically dominated
by a single age range, and the youngest population located in the inner edge of the arm maps the location of the chain of complexes observed 
in this arm.

Upcoming work is applying this method to the rest of the sample of 6 galaxies from the original work, Paper I. To analyze with more detail now the age patterns.
Next, we are studying also the more complex case and model of dependence between pixels of the image.
To develop a parallelization of the algorithm, high perfomance computing to emcee, and / or HMC (programming in Python and R) to get better computing times. 
And finally, to get a wider, new sample of galaxies, to apply the method and determine not only their age maps, but also to study possible patterns/gradients in the metallicity maps. 

%%%%%%%%%%%%%%%%%%%%%%%%%%%%%%%%%%%%%%%%%%%%%%%%%%
%%%%%%%%%%%%%%%%%%%%%%%%%%%%%%%%%%%%%%%%%%%%%%%%
%\newpage 
\appendix

\section{Empirical Model: Starbust 99}\label{sec:empiricalSB99}

Comparison of UV and \Ha\ fluxes provides a good indication of the recent star formation history. 
As a young star forming region evolves, the \Ha\ emission line declines before the UV emission does, leading to a decrease in $F_{H\alpha}/F_{FUV}$, which is very sensitive to the age of the youngest stellar population. 
Images at these wavelengths can be examined to track the recent star formation history and distribution, and to analyze how the SFH correlate with the galactic structure and the galactic environment.

Model \Ha/FUV flux ratios were generated from Starburst99, as a function of age, to be compared with the measured flux ratios at each pixel in the image. 
The full models cover the age range $10^6$ to $10^9$ yr  in steps of 1~Myr with spectral 
energy distributions  (SEDs) spanning 100 \AA\ to 1 $\mu$m in wavelength. However, we restricted the 
ages from 1 to 10 Myr as we are only interested in the youngest stellar populations responsible for the \Ha\ and UV output.

The code can be run with two different star formation modes:  an instantaneous burst or continuous. It has three alternatives in the stellar initial mass function, (IMF):
a Salpeter law with $\alpha=2.35$ and $M_{up} = 100M_{\odot}$; a truncated Salpeter law with $\alpha=2.35$ and $M_{up} = 30M_{\odot}$; and a Miller-Scalo law with $\alpha=3.3$ and $M_{up} = 100M_{\odot}$.
And five metallicities are available for each IMF:   $Z = $ 0.04, 0.02 (solar, $Z_{\odot}$), 0.008, 0.004 and 0.001. 

We take as our reference model a Salpeter IMF, the interpolated metallicity value {Z=0.02 (according to  \cite{Miller1996}, M74 has 12+log(O/H)$\sim$8.15)}, and an instantaneous star formation law, since it is more sensitive to the age variation for younger regions (see paper I).

The age map is contoured in four age bins from the set 0 -- 4 Myr, 4 -- 6 Myr, 6 -- 9 Myr and older than 9 Myr, (this binning corresponds to the internal precision of the flux ratio, Paper I) see right panel of Fig.\ref{fig1} . 
We find in the resulting age map that the youngest star forming population (0 -- 4 Myr) surrounds an older one ($>$9 Myr) located at the centre of the complex,  
and a clear center to outer rim age gradient. 

The $F_{H\alpha}/F_{FUV}$ depends on the model parameters, such as the IMF, the metallicity, the SFH, etc.  
To assess the suitability and robustness of our reference model we calculated the 
effect of changing various model inputs on the SB99 $F_{H\alpha}/{F_{FUV}}$ ratios (cf. Fig.2 of Paper I).
%%%
The photometric uncertainty on the F$_{H\alpha}/$F$_{FUV}$ ratio also affects to the age calibration.
Aside the model and photometric uncertainties, the age-dating technique is subject to a number of 
potential sources of systematic error as: 
(i) the lowest limit on cluster mass allowable for our assumptions on ionizing flux, (ii) the effect
of changing the spatial bin size, and (iii) the effect of changes to the metallicity and IMF assumptions
in the model.
We analyze the reliability and robustness of our age-dating technique taking into account all these uncertainties. See Paper I for further details. 

The total uncertainty of the extinction corrected \Ha/FUV flux ratio
is estimated, by error propagation, from the \Ha\ , FUV and TIR flux 
uncertainties as:
$\Delta log(F_{H\alpha}/F_{FUV})_0  = \Delta [0.4(A_{H\alpha} - 
A_{FUV})] + \Delta log(F_{H\alpha}/F_{FUV}) = \frac{1}{ln10} [f(IRX)
(E_{TIR}+E_{FUV}) + (E_{H\alpha}+E_{FUV})]$.
Where $f(IRX)=4/35\mid-0.0999IRX^2+0.7044IRX+0.196\mid$, and 
$IRX=log(F_{TIR}/F_{FUV})$ (from equation of  \cite{Buat2005}). 
These uncertainties are computed pixel by pixel in the image. An average 
of the relative errors of the respective images are $E_{H\alpha}
\simeq5\%$, $E_{FUV}\simeq25\%$, and $E_{TIR}\simeq6\%$, resulting in an 
overall uncertainty in the F$_{H\alpha}/$F$_{FUV}$ flux ratio of 
$\sim$28\%.

Now, we calculate the underlying stellar mass per pixel and compare with 
the values of the lower mass limit $M^{min}$, given by  \cite{Cervi2003}.
The pixel stellar mass is a lower limit derived by comparing the 
linearly scaled extinction-corrected observed L$_{FUV}$ with the highest 
expected value from SB99, for a stellar population mass of $10^6$ 
M$_{\odot}$ at the youngest cluster age (1 Myr). In our map, these 
lowest mass values range 
log$M^{min}=3.2-5.2$, with an average of $3.84\pm0.33$, around the 
$M^{min}$ estimated by  \cite{Cervi2003}. 
Binning the IC~2574 image by $3\times3$ puts the spatial sampling of 
IC~2574 well above the threshold of  $M^{min}$ (cf. Fig.5 of Paper I ), 
to remove any cause of bias by IMF fluctuations.

%%%%%%%%%%%%%%%%%%%%%%%%%%%%%%%%%%%%%%%%%%%%%%%%

%%%%%%%%%%%%%%%%%%%%%%%%%%%%%%%%%%%%%%%%%%%%%%%%

%\section*{acknowledgment}

%%%%%%%%%%%%%%%%%%%%%%%%%%%%%%%%%%%%%%%%%%%%%%%%
%%%%%%%%%%%%%%%%%%%%%%%%%%%%%%%%%%%%%%%%%%%%%%%%

\end{document}